\def\ang{\stackrel{\rm o}{\rm A}\/}
\def\Rho{\widetilde{\rho}\/}
\def\K{{\bf K }\/}

\def\r{{\bf r }\/}
\def\k{{\bf k }\/}
\def\n{\noindent}

\documentclass{elsart}
\usepackage{epsfig}
\usepackage{graphicx}
\usepackage{amssymb}

\begin{document}

\begin{frontmatter}

\title{A local-density approximation for the exchange energy functional for excited
states  : the band gap problem.}
\author[label1,label2]{Moshiour Rahaman}
\author[label1,label2]{Shreemoyee Ganguly}
\author[label3]{Prasanjit Samal}
\author[label2]{Manoj Kumar Harbola}
\author[label1,label4]{Tanusri Saha-Dasgupta}
\author[label1,label2,label4]{Abhijit Mookerjee}
\address[label1]{Department of Materials Science,  S N Bose National Centre for Basic Sciences,Block JD, Sector III, Salt Lake, Kolkata 700098,India}
\address[label2]{Department of Physics, Indian Institute of Technology, Kanpur 208016, India.}
\address[label3]{Department of Physics, National Institute of Science Education \& Research, 
Institute of Physics, Sachivalaya Marg, Bhubaneswar  751005, India.}
\address[label4]{Advanced Materials Research Unit, S N Bose National Centre for Basic Sciences,Block JD, Sector III, Salt Lake, Kolkata 700098, India}

\begin{abstract}
 We present excited states density functional theory (DFT) to calculate band gap for semiconductors and insulators. For the excited states exchange-correlation functional, we use a simple local density approximation (LDA) like functional and it gives the result which is very closed to experimental results. The linear muffin-tin potential is used to solve the self consistent Kohn-Sham equation.
\end{abstract}

\end{frontmatter}


\section{Introduction}
First-principles calculations based on the density functional theory (DFT) have been
eminently successful for the study of ground state properties of electrons in a 
solid \cite{dft1}. However, it has always been understood that the spectrum
and wave-functions of the associated one-electron like Kohn-Sham equation have no
specific significance beyond the fact that they are used to obtain the ground-state
density \cite{ks}. Thus the unoccupied part of the Kohn-Sham spectrum cannot be expected
to describe correctly the excitations of a many-electron system. On the other hand,
the energy eigenvalue of the highest occupied orbital, i.e. the orbital energy
corresponding to the top of the valence band, is the exact ionization
potential of the system \cite{pplb}.  This is known as the ionization potential theorem.  Since
adding an extra electron at the bottom of the conduction band would hardly change
the density of a bulk system, the ionization potential theorem leads one to expect
that the difference in the Kohn-Sham orbital energies corresponding to the bottom
of the conduction band and the top of the valence band should give the correct 
band gap of semiconductors and insulators.  However it is well known that such
an interpretation leads to gross underestimatation of the band gaps of semiconductors
in comparison to the experiments. This is true even in solids like Ge, Si and the 
III-V semi-conductors which are not ``strongly correlated". This underestimation 
of the band gap has been explained in terms of the derivative discontinuity
 \cite{lp1,shamsch1} of the exchange-correlation energy functional. To circumvent
this difficulty, we propose to look at the calculation of the band gap as an 
excited-state problem. As such, the question we ask is: could we set up a DFT for 
excited states, calculate the total energy for appropriate excited-state and obtain
the excitation energy as its difference from the ground state total energy. We may 
then go on to interpret the band gap in a semi-conductor as the difference of the 
energies between the ground state and an excited state where an electron in the 
highest occupied (HO) state of the ground state is removed from it and excited to the 
lowest unoccupied (LU) state. 

The mathematical basis of a density functional theory for total energies in excited
states had been set up by G\"orling \cite{gor} and Levy and Nagy \cite{ln}, based 
on the constrained search approach \cite{cs}.  The theory has been put on a strong
footing by Samal and Harbola \cite{sh1}. The energy of the excited state of 
electrons in a solid may be expressed as a bi-functional of both the excited state
density and the ground state density.
 
\begin{equation} 
E[\rho_{\rm ex},\rho_{\rm gr}]\ =\ F[\rho_{\rm ex},\rho_{\rm gr}]\ +\ \int\ d^3\r\ V(\r)\rho_{\rm ex}(\r) 
\label{eq1}
\end{equation}

\noindent where $V(\r)$ is the ion-electron potential. $F[\rho_{\rm ex},\rho_{\rm gr}]$ is
a bi-functional of the excited state density  
$\rho_{\rm ex}(\bf r)$  and the ground-state density $\rho_{\rm gr}(\r)$.
It is defined \cite{ln} from the constrained-search formulation :

\begin{equation} 
F[\rho_{\rm ex},\rho_{\rm gr}]\ =\ \stackrel{\displaystyle\mathrm{min}}{\phantom{x}_{\Phi\rightarrow\rho_{ex}}} \langle\Phi\vert(\hat{T}+\hat{V}_{ee})\vert \Phi\rangle 
\label{eq2}
\end{equation}

\noindent where the constrained search is carried out with the wavefunctions giving the excited-state 
density;  
 $\hat{T}$ and $\hat{V}_{ee}$ are the kinetic energy and electron-electron
interaction potential operators respectively  and the wavefunctions $\Phi$ are orthogonal to all the lower states determined by 
$\rho_{\rm gr}$. For example, in defining $F[\rho_{\rm ex},\rho_{\rm gr}]$ for the first
excited state, the $\Phi$-s chosen would be such that they are orthogonal to the ground state wave-function obtained from $\rho_{\rm gr}$. The exchange-correlation energy functional is then :

\begin{equation}
 E_{\rm xc}[\rho_{\rm ex},\rho_{\rm gr}] \ =\ F[\rho_{\rm ex},\rho_{\rm gr}]-T_0[\rho_{\rm ex}] - V_{\rm H}[\rho_{\rm ex}] 
\label{eq3}
\end{equation}

\noindent Here $T_0$ represents the kinetic energy of a  non-interacting
system of electrons, suitably chosen so that its charge density is the same as that of the
interacting system. The effective potential seen by the system of non-interacting
electrons is variationally obtained and the single-particle wave-functions of the non-interacting system satisfies the Kohn-Sham equation :

\begin{equation}
\left[ -\frac{1}{2}\nabla^2 + v(\r) + \int d^3\r'\ \frac{\rho_{\rm ex}(\r')}{\vert\r-\r^\prime\vert} + v_{\rm xc}^{\rm ex}(\r)
\right]\phi_i(\r)
\  =\ \varepsilon_i\ \phi_i(\r) 
\label{eq4}
\end{equation}

\noindent where $V(\r)\ =\ \sum_n v(\r-{\bf R}_n)$ and {\bf R}$_n$ are the positions of the ion-core labelled by $n$. 
The excited state density is obtained from : 
$\rho(\r)\ =\ \sum_i\ f_i\ \vert\phi_i(\r)\vert^2 $ where
the occupation number $f_i$ is 0 or 1 depending on whether the state labelled by 
$i$ is occupied or not.

As in the ground-state theory, the exchange-correlation energy functional in
excited-state theory too has to be approximated. The problem of finding an
excited-state exchange-correlation energy functional is a challenging one and 
is further compounded by the fact that the functional is  expected to be 
state dependent.  Thus different functionals may have to be constructed for different
classes of the excited-states.  Recently, Samal and Harbola \cite{sh2} have 
proposed a local-density approximation (LDA) for the exchange energy functional
for a theory of excited-states with one gap. The functional has
been applied to a large number of atomic excited-states to obtain very accurate
excitation energies for single and double-electron excitations. The functional
has precisely the form required to calculate band-gaps of semiconductors using
excited-state DFT. The purpose of the present work is then to employ this excited-state
DFT to obtain the band-gaps of a number of semiconducting and insulating materials.
In the following we give a brief description of the methodology employed by
us and then present our results.

As we discussed earlier, the band gap in a semiconductor or an
insulator can be thought of as the lowest energy required to excite one electron 
out of a many-electron ground state to one of the excited states of the system.  
Thus, in terms of the transition energy,

\begin{equation}
\Delta E =  E_{\rm ex}[\rho_{\rm ex},\rho_{\rm gr}]-E_{\rm gr}[\rho_{\rm gr}]
\label{eq5}
\end{equation} 

\noindent where $E_{\rm ex}[\rho_{\rm ex},\rho_{\rm gr}]$ is the total excited-state energy functional of (1) minimized by varying $\rho_{\rm ex}$ and 
$E_{\rm gr}[\rho_{\rm gr}]$ is the ground-state energy corresponding to the configurations
mentioned above.

To obtain the total energies we turn to the solution of the  variationally obtained 
Kohn-Sham equation (\ref{eq4}) \cite{ks}.  
 The charge density is given by,
\[
\rho(\r)=\sum_{i=1}^{\infty}f_{i}\ |\varphi_{i}(\r)|^{2}\quad \mathrm{with}\quad 
\sum_{i=1}^{\infty}\ f_{i}=N_e \]

\noindent Here $\rho$ is either the ground state or the excited state densities depending upon the occupation numbers $f_i$. The total energy functional is  :

\begin{eqnarray}
E[\rho] & = &\sum_{i=1}^{\infty}\ f_{i}\ \epsilon_{i}-\frac{1}{2}\int\int d^3\r\ d^3\r'\ \frac{\rho(\r)\rho(\r^\prime)}
{|\r-\r^\prime|} 
 -\int d^3\r\ \frac{\delta \widetilde{E}_{\rm xc}}{\delta\rho(\r)}\ \rho(\r)\ +\ \widetilde{E}_{\rm xc} 
\nonumber\\ & =&  E_{\rm KS}- E_{\rm H}[\rho]+ \widehat{E}_{\rm xc}  
\label{eq8}
\end{eqnarray}

\noindent Here $\rho$ is either $\rho_{\rm gr}$ or $\rho_{\rm ex}$ depending on whether we are talking about the ground or the excited state; $\widetilde{E}_{\rm xc}$ is the ``modified exchange" energy functional which includes the difference between
the interacting and non-interacting kinetic energies as well. This is of the 
form $\widetilde{E}_{\rm xc}[\rho_{\rm gr}]$ or $\widetilde{E}_{\rm xc}[\rho_{\rm ex},\rho_{\rm gr}]$ , again 
depending upon whether we are talking about the ground or the excited state. $E_{\rm H}$ is the Hartree energy
and $\widehat{E}_{\rm xc}$ is the contribution from the modified exchange-correlation energy functional given by the last two terms in the Eq. (\ref{eq8}). Note that for the sake of notational clarity the
bi-functional notation for the excited state functionals in the above equation has been dropped.

\section{The excited state exchange-correlation functional}

There have been many attempts at obtaining accurate but approximate forms for the
exchange-correlation functional for the ground state. The simplest of them, the
LDA, follows the Dirac \cite{dirac} approach and proposes the functional on the
basis of exchange energy for the homogeneous electron-gas. As mentioned above,
adopting the same approach, Samal and Harbola \cite{sh2} proposed an excited-state
functional for the exchange-energy and applied it to atoms to obtain accurate
excited-state energies.  
In the following we discuss the construction of such a functional for
semi-conductors and insulators and employ it to get the band gap
of these materials as defined by Eq.(\ref{eq5}) above. The accuracy or otherwise of our 
proposition will be evident from the applications to several semi-conductors and 
insulators. 

\begin{figure} 
\centering
\includegraphics[width=3in,height=2in]{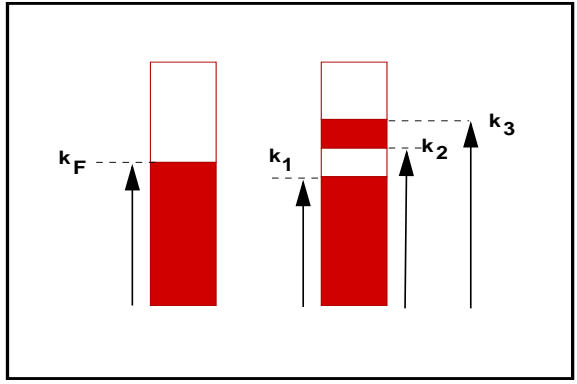}
\caption{The pictures of the ground state configuration of a HEG (left)
and an excited state (right).}
\label{fig1}
\end{figure}

The standard procedure for setting up of the exchange-correlation functional is to start from the equivalent expression for the homogeneous electron 
gas (HEG).  This is shown in Fig. \ref{fig1}. In the semi-conductor or insulator  in its ground state, the 
electrons are filled up to the Fermi level, that is, in reciprocal space electrons 
occupy states labelled by wave vectors from $k=0$ to $k_{F}$. The same picture is 
true for the HEG, but in addition we have 
\begin{equation}
 k_F^3\ =\ 3\pi^{2}\rho_{\rm gr}(\r)
\label{eq9}
\end{equation} 
where $\rho_{\rm gr}(\r)$ is the almost uniform electron density in the ground state. 

On the other hand, for the lowest excited state in the semi-conductor or insulator, 
the electron in the HO state  in the ground state configuration 
 is removed from that state and now occupies the lowest unoccupied state.
  For the HEG this excited state configuration  corresponds
to the following : first, we shall call all states which were occupied in the ground state as well
as in the excited state : {\sl core states}. If $\rho_{\rm core}(\r)$  is the
electron density due to the electrons occupying the core states, then all states 
from $k\ =\ 0$ to $k_1$ are occupied, where 

\begin{equation}
 k_1^3 \ =\ 3\pi^2\ \rho_{\rm core}(\r)
\label{eq10}
\end{equation}
Next, if $\rho_{rem}(\r)$ is the electron density due to the electron 
in the HO state of 
ground state configuration which has been removed in the excited state configuration, then 

\begin{equation}
 k_2^3\ -\ k_1^3\ =\ 3\pi^2\ \rho_{\rm rem}(\r)
\label{eq11}
\end{equation}

Finally, if $\rho_{add}$ is the electron density due to the electron in the lowest excited state of the ground state configuration, 
 which has been added in the excited state configuration, then

\begin{equation}
 k_3^3\ -\ k_2^3\ =\ 3\pi^2\ \rho_{\rm add}(\r)
\label{eq12}
\end{equation}

All states between the labels $k_1$ and $k_2$ which were occupied in the ground state become
unoccupied in the excited state. Whereas all states between the labels $k_2$ and $k_3$ 
which were unoccupied in the ground state become occupied in the excited state.

The ground state density is $\rho_{\rm gr}(\r)\ =\ \rho_{\rm core}(\r)\ +\ \rho_{\rm rem}(\r)$ and the excited
state density is $\rho_{\rm ex}(\r)\ =\ \rho_{\rm core}(\r) + \rho_{\rm add}(\r)\ =\ \rho_{\rm gr}(\r)-\rho_{\rm rem}(\r)+\rho_{\rm add}(\r)$.
Note that in an excited state of the system the electrons will occupy the $k$-space differently from the ground state.
\vskip 0.2cm
As stated before, the standard starting point for building up  the ground state exchange-correlation functional,
whose exact form is in general unknown, is the exact expression for the HEG. We shall use the
same idea for the excited state. This new functional will be the basis for our LDA generalized for
excited states (MLDA).

The exchange energy in the MLDA can be obtained by integration over the reciprocal space equi-energetic spherical shell surfaces. In a recent paper Samal and Harbola \cite{sh2} have explicitly
obtained the expression for the exchange energy functional for the excited state, as described above. We shall quote their results here and refer the interested reader to the above mentioned reference for the mathematical details. 

\[ E_x^{MLDA} = E_x^{\rm core} + E_x^{\rm add} + E_x^{\rm add-core} \]

\noindent where

\begin{eqnarray}
 E_{x}^{\rm core}  & = & -\frac{1}{4\pi^3}\int d^3\r\ k_1^4 \nonumber\\
 E_{x}^{\rm add} & = & -\frac{1}{8\pi^3}\int d^3\r\ \left[2(k_3^3-k_2^3)(k_3-k_2)+(k_3^2-k_2^2)^2\ln\left(\frac{k_3+k_2}{k_3-k_1}\right)\rule{0mm}{4mm}\right]
  \nonumber \\
 E_{x}^{\rm add-core} & =&  -\frac{1}{8\pi^3}\int d^3\r\ \left[\rule{0mm}{4mm}2(k_3-k_2)k_1^3+2(k_1^3-k_2^3)k_1 + \ldots 
\right.\nonumber\\ & &  \left.(k_2^2-k_1^2)^2\ln\left(\frac{k_2+k_1}{k_2-k_1}\right)
 -(k_3^2-k_1^2)^2\ln\left(\frac{k_3+k_1}{k_3-k_1}\right)\right] 
\label{eq13}
\end{eqnarray}

\noindent It is easy to check from above that when $k_1=k_2=k_2=k_F$,  Eq. (\ref{eq13}) reduces 
to the ground state exchange functional.  Equations (\ref{eq10})-(\ref{eq13}) will then provide the 
basis for the calculation of the total energies. If we now go back to Eq. (\ref{eq8}), we note that the expressions for the ground state and excited state
total energies are :

\begin{eqnarray*}
E_{\rm gr} & = & E^{\rm KS}_{\rm gr} - E_{\rm H}[\rho_{\rm gr}] + \widehat{E}^{LDA}_{x}[\rho_{\rm gr}]\\
E_{\rm ex} & = & E^{\rm KS}_{\rm ex} - E_{\rm H}[\rho_{\rm ex}] + \widehat{E}^{MLDA}_{x}[\rho_{\rm ex},\rho_{\rm gr}]\\
\end{eqnarray*}

\noindent Here $E^{\rm KS}$ refers to $\sum_i f_i \epsilon_i$ where $f_i$ are the occupation
numbers already defined and $\epsilon_i$ are the Kohn-Sham energies. 
Assuming no change in the 
orbital eigenvalues corresponding to the ground- and excited-state configurations, 
which is a good assumption for bulk systems, the band gap will be given by the minimum
value for different excitations of :

\begin{eqnarray}
\Delta E &=& \Delta E^{\rm KS} - \Delta E_{\rm H} + \left\{\widehat{E}^{MLDA}_x[\rho_{\rm ex};\rho_{\rm gr}]-\widehat{E}^{LDA}_x[\rho_{\rm gr}]\right\}\nonumber\\
\label{eq14}
\end{eqnarray}

\noindent where $\Delta E^{\rm KS}$ is the conventional Kohn-Sham gap, $\Delta E_{\rm H}$ is the
difference in the Hartree energy corresponding to the excited-state and the
ground-state densities, where both $\rho_{\rm ex}$ and $\rho_{\rm gr}$ are calculated from the
orbitals obtained from a single ground-state calculation. If the sum of the last
two terms is small, we may expect the Kohn-Sham gap to be the band-gap. However, 
we expect the major correction to the conventional Kohn-Sham gap 
to come from these terms. What these terms accomplish is to replace the LDA 
exchange energy corresponding to the excited-state density $\rho_{\rm ex}$ by the MLDA 
exchange energy.  The latter is tailor-made - and therefore more appropriate - for 
getting exchange energy for an excited-state of the kind considered for the band 
gap calculation. We note that the functional proposed by Samal and Harbola also has
self-interaction correction (SIC) terms in it. For extended systems we expect these terms
to be negligible and shall justify our assumption by application to a few
of the systems studied by us.

\section{Application to Solids}

We shall base our calculations in solids on the tight-binding linear muffin-tin 
orbitals method within the atomic-sphere approximation (TB-LMTO-ASA) 
 \cite{ak1}-\cite{ad}. In this method the muffin-tin spheres are inflated to the 
Wigner-Seitz cell volume and the interstitial contribution is neglected. 
The wave-function is expanded in a basis of linearized muffin-tin orbitals :

\begin{eqnarray*}
 \psi_\sigma(\r,E_{\k n\sigma})& =& \sum_{RL} u_{RL\sigma}(E_{\k n\sigma})\ \chi_{L\sigma}^R(r,E_{\k n\sigma})\
Y_L(\hat{r})\ \xi_\sigma \\
\chi_{L\sigma}^R(r, E_{\k n\sigma}) & = & \left[\ \varphi_{R\ell\sigma}(r)+\left(E_{\k n\sigma}-E_{\nu R\ell})\ \dot{\varphi}_{R\ell\sigma}(r)\right)\right]
\end{eqnarray*}

Here, $Y_L(\hat{r})$ are the spherical harmonics and $\xi_\sigma$ are the spinor wave-functions. 
The muffin-tin orbitals  $\chi_{L\sigma}^{R}(r,E_{\k n\sigma})$ 
belong to an atom   of type $R$, the angular momentum L=($\ell,m$), 
$\varphi_{R\ell\sigma}(r)$ are the partial wave solutions in the muffin-tin spheres,
$\dot{\varphi}_{R\ell\sigma}$ are the energy derivatives of the partial waves, both
evaluated at the expansion energies $E_{\nu R\ell\sigma}$ used for linearization. Tail 
cancellation ensures that the muffin-tin orbitals labelled by $R$ vanish outside the atomic
spehers centered at $R$. The quantum states of the electrons in the solid are labelled by
the quantum numbers $\K = (\k n\sigma)$.

From the Kohn-Sham band-structure we can identify the HO and LU states in reciprocal space. Using
these values of $\K$ we construct the removed and added electron densities. The core electron
density is obtained from the atomic-like the core states.
These expressions are then used in Equations (\ref{eq10})-(\ref{eq14})
 to obtain the band gap. We again point out
that both the ground- and the excited-state densities are constructed from a single
ground-state calculation.  This is justified since one does not expect lowest unoccupied state to
change much even if an excited-state KS calculation is performed.  An advantage
of this is that the computational demand of our calculations is pretty much at the
same level as the ground-state calculations.

Application of our methodology to molecules and clusters have already been carried out 
without any difficulty earlier \cite{sh1}-\cite{sh2}. However, in its application to
solids there have been two points of confusion. This is related to the fact that 
 in a bulk solid the extended Bloch-like electronic wave-functions for a given quantum
state are infinitely thinly spread over the whole solid. The charge densities
corresponding to a single quantum state are also $O$(1) while the charge density of the
valence cloud is $O$(N) ($N$ is the number of atoms)\footnote{A quantity is called $O(N^\alpha)$
if it scales as $N^{\alpha}$ with the number $N$  of atomic-spheres in the system. For example, total
energy is $O(N)$ while the energy per atom is $O(1)$.}.

There are two questions we must answer : First, are we really exciting a {\sl single} electron out of a 
sea of $N$ ($N\rightarrow\infty$) electrons, from the 
HO  to the LU state ?  Secondly, 
can we use this infinitely small ($O$(1) as compared to $O$(N)) change to produce the finite energy
differences which we define to be the band gap in Eq. (\ref{eq14}) ?

In this section we shall argue that the way we obtain the energy differences within the LMTO technique by normalizing
the solution within a Wigner-Seitz cell and considering only energies per cell addresses this problem successfully.
We should also note that the arguments given is not confined to the LMTO, but holds good for any electronic structure
method that are based on ``cellular" techniques, e.g. KKR, APW or LAPW. 

As we have seen earlier, quantum states of electrons in a solid are labeled by $\K$. If there are $N\ (N\rightarrow \infty)$ atoms constituting the solid each contributing $m$ electrons to the valence electron cloud, 
then the total number $mN$ of available states
are occupied, each state by one electron. If $\rho(\r)$ is the total charge density of occupied electron states, then 

\[ \int_\Omega\ d^3\r\ \rho(\r)\ =\ mN \quad \mbox{integrated over the whole volume.}\]

In all electronic structure techniques which divide the solid into identical
Wigner-Seitz cells (WSC) centered at the sites $R$ such that :
\[ \Omega = \bigcup_{R=1}^{N}\ \Omega_R\quad \mbox{where}\quad \Omega_R\bigcap\Omega_{R'} = \emptyset\quad \forall R'\neq R \]

\noindent where $\Omega$ is the volume of the solid and $\Omega_R$ that of a WSC centered at $R$,   the basis of representation are WSC centered functions of the form :

\begin{equation}
 \chi_{L\sigma}^R(\r)\ =\ \left\{\begin{array}{ll}
                     \chi_{L\sigma}^R(r)Y_L(\hat{r})\xi_\sigma & \mathrm{for\ \r\in\ WSC\  centered\ at\ } R\\
                     0 & \mathrm{otherwise}
                     \end{array}\right.
\label{A1}
\end{equation} 

\n $L$ is a composite index, which for the KKR or the APW and their linearized versions, LMTO and LAPW, denote
the angular momenta indices $(\ell m\sigma)$. These basis functions are centered at WSCs and
do not overlap. The charge density also may be broken up as :

\begin{eqnarray}
 \rho(\r) & = & \sum_{R=1}^N\ \sum_{\K}\ f_{\K}\ \rho_{\K}^R (\r)= \sum_{R=1}^N\  \rho^R(\r) 
 \\
 \mathrm{where}\ && \rho^R(\r)=  \left\{\begin{array}{ll}
                     \rho^R(\r) & \mathrm{for\ \r\in\ WSC\  centered\ at\ } R \\
                     0 & \mathrm{otherwise}
                     \end{array}\right.\nonumber
\end{eqnarray}

Note that each quantum state labelled by \K can be occupied by a single electron. 
Here $f_{\K}$ is the occupation number of the quantum state labelled by $\K$. For example, in the ground state 
$mN$ of the lowest energy states are occupied, so $mN$ of the $f_\K$ are 1 the rest are 0. 
In a crystalline solid, the densities within each  WSC labelled by $R$ are identical. However, 
since these partial charge densities have the same structure as Eq. (\ref{A1}), and are 0 outside
the WSC labeled by $R$, we shall retain this label to indicate this. From the above it is clear that,
since $\rho(\r)$ integrates to $mN$ over $N$ WSC : 

\begin{equation}
\int_{\Omega_R} \ d^3\r\ \rho^R(\r) = m \quad \mbox{and} \quad 
\int_{\Omega_R} \ d^3\r\ \rho_{\K}^R(\r) = 1/N
 \label{A2}
\end{equation}

In the output of the cellular methods (including the KKR, APW, LMTO,LAPW), the charge densities per WSC are scaled such
that they are normalized to 1 within a cell.

 \[ \int_{\Omega_R} \ d^3\r\ \widetilde{\rho}_{\K}^R(\r) = 1 \]

 So that the {\sl scaled} charge densities per $\K$ labelled state which we get as output from the cellular methods are related to the true charge densities per state as~:

\begin{equation}
\rho_{\K}^R(\r) = (1/N) \widetilde{\rho}_{\K}^R(\r) 
\label{A3}
 \end{equation}

If we now remove or add a scaled charge density $\widetilde{\rho}_{\K}^R(\r)$ integrating to 1 in a WSC, then
that is equivalent to removing or adding a true charge density integrating to 1/N in that cell. The total charge
density added or removed in the whole solid then integrates over the whole volume to 1. In other words, we {\sl are}
exciting 1 electron from the valence cloud in the solid from the HO to LU state.  

The total energy expressions of Eq.(\ref{eq8}) can also be broken up as :

\[ E[\rho]\ =\ \sum_{R=1}^N\ E_R[\rho^R(\r)] \]

We now go back to the way we construct the charge density of the excited state : We divide the
total charge density of the ground state into a density of the `core' which does not change on
excitation $\rho_{\rm core}(\r)$ and that due to the HO state $\rho_{\rm rem}(\r)$. We then remove $\rho_{\rm rem}(\r)$
and replace it by the density due to the LU $\rho_{\rm add}(\r)$.

\begin{eqnarray*}
 \rho_{\rm ex}(\r)  & = & \rho_{\rm core}(\r)+\rho_{\rm add}(\r) = \rho_{\rm core}(\r)+\rho_{\rm rem}(\r)+(\rho_{\rm add}(\r)-\rho_{\rm rem}(\r))\\
&  =& \rho_{\rm gr}(\r)+\delta\rho (\r)
 \end{eqnarray*}

We also note from Eq. (11) that  when $\delta\rho\rightarrow 0$ $E^{MLDA}[\rho_{\rm gr}]=E^{LDA}[\rho_{\rm gr}]$. Combining these two, we get~:

\begin{eqnarray}
\Delta E[\rho] &\ =\ &E^{MLDA}[\rho_{\rm ex}(\r)]-E^{LDA}[\rho_{\rm gr}(\r)]\nonumber\\
 &\  =\ & E^{MLDA}[\rho_{\rm gr}(\r)+\delta\rho (\r)]-E^{LDA}[\rho_{\rm gr}(\r)]\nonumber\\
 &\ =\ &\sum_{R=1}^{N} \left\{
\rule{0mm}{4mm} E^{MLDA}_R[\rho^R_{\rm gr}(\r)+\delta\rho^R(\r)]-E^{LDA}_R[\rho^R_{\rm gr}(\r)
]\right\}\nonumber\\ 
& \ =\ & \sum_{R=1}^N \ \left\{ \int_{\Omega_R} d^3\r\ \delta\rho^R(\r) \frac{\delta E_R[\rho^R(\r)]}{\delta\rho^R(\r)}\left.\rule{0mm}{7mm}\right\vert_{\rho^R=\rho_{gr}^R} +\ldots\right. \nonumber\\
& & \left.+ \frac{1}{2!} \int_{\Omega_R}\int_{\Omega_R} d^3\r d^3\r^\prime
\ \delta\rho^R(\r)\delta\rho^R(\r')\ \frac{\delta^2 E_R[\rho(\r)]}{\delta\rho^R(\r)\delta\rho^R(\r ')}\left.\rule{0mm}{7mm}\right\vert_{\rho^R=\rho_{gr}^R}+\ldots \right\} \nonumber\\
&\ =\ & \sum_{R=1}^N \ \left\{\int_{\Omega_R}d^3\r\ \delta\rho^R(\r)\ A[\rho_{\rm gr}^R(\r)] + o(\{\delta\rho^R(\r)\}^2)\right\} \nonumber\\
&\ =\ & N \left\{\int_{\Omega_R}d^3\r\ \delta\rho^R(\r)\ A[\rho_{\rm gr}^R(\r)] + o(\{\delta\rho^R(\r)\}^2)\right\}\nonumber\\ 
&\  =\  & \int_{\Omega_R} \d^3\r \ \delta\widetilde{\rho}^R(\r)\ A[\rho_{\rm gr}^R(\r)] + N\ o(\{\delta\rho^R(\r)\}^2)
\label{A10}
\end{eqnarray}

Here $o(x^n)$ stands for terms which scale as $x^m, m\geq n$. Note that $\delta\rho^R(\r)\sim 1/N$ so that all terms
 $N \ o\left(\{\delta\rho^R(\r)\}^2\right)$  become negligible as $N\rightarrow\infty$. In this limit we may
forget these latter terms.

The last but one step follows from the fact that all WSCs are identical in a crystalline solid. 
We note from Eq.(\ref{A2}) that although the total energies on the right side of the above
equation are individually $O(N)$, the difference $\Delta E$ should be $\sim  O(1)$. Since $\delta\rho(\r)$
scales like $1/N$ and the summation of $R$ provides a factor of $N$, the functional $A[\rho_{\rm gr}^R(\r)]$
$\sim\ O(1)$.

The energies calculated in the TB-LMTO are energies {\sl per unit cell}. The energy difference calculated within the TB-LMTO is :
\begin{equation}
\Delta E[\Rho^R]\ =\  
  \int_{\Omega_R} \delta\Rho^R(\r)\ A[\Rho_{\rm gr}^R(\r)] 
\label{A11}
\end{equation}

The only way in which both $\Delta E[\rho]$ and $\Delta E[\Rho^R]$ can be $O$(1) is if the
functional $A[\rho]$ has a specific scaling behaviour with $\rho$. Let $A[\Rho]=A[N\rho]=N^\alpha A[\rho]$,
then $\alpha = 0$ for the above to hold. In that case,
it now follows that :

\[ \Delta E[\Rho^R] \ =\ \Delta E[\rho] \ =\ E_g \]

The calculation in the solid is thus reduced to the calculation within a single WSC using the scaled 
charge density $\Rho(\r)$ and the differences in energies per unit cell. We are exciting one electron from
the entire valence cloud of the solid from the HO to LU state and the total energy difference is  E$_g$ = $O$(1).

\section{Results and Discussion}
\begin{figure}[th!] 
\centering
\includegraphics[width=3in,height=2in]{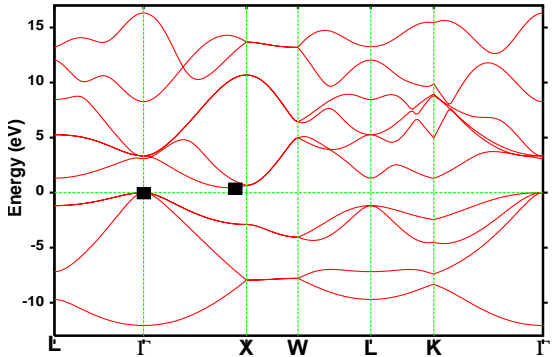}
\caption{The Kohn-Sham band structure of Si showing the HO and
LU  states used in our calculations}
\label{fig2}
\end{figure}

\begin{figure}[bh!] 
\centering
\rotatebox{270}{\includegraphics[width=3.in,height=4.5in]{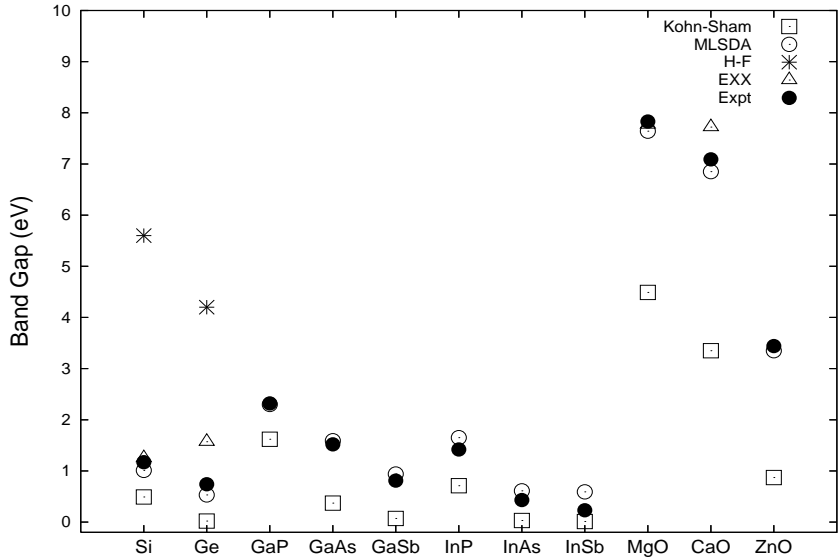}}
\caption{Band gaps calculated by different methods compared with the
experimental band gaps of a series of semi-conductors and insulators.}
\label{gap}
\end{figure}

We shall first illustrate our calculations with a specific example, i.e. that of Si.
The Fig. \ref{fig2} shows the Kohn-Sham band structure of Si and the HO and LU states as
obtained from the regular Kohn-Sham LDA calculation. In addition to these states,
the calculations also give the Kohn-Sham gap $\Delta E^{\rm KS}$. 
These states are then used to construct the removed and added electron densities 
for the excited state. When employed in Eq. (\ref{eq14}), these densities yield the band
gap of silicon.  The numerical results are shown in Table 1 while a comparison is shown in Fig \ref{gap}.  

As is clear from the Table 1 and Fig. \ref{gap}, our
method improves the Kohn-Sham band gap substantially, bringing it very close to
the exact exchange \cite{kotani1} and experimental \cite{cox2} band gaps.
Interestingly the correction to the Kohn-Sham band gap is 0.52 eV and arises essentially
from the exchange-energy difference between the ground- and the excited-state.  
This difference is very close to the derivative discontinuity of 0.58 eV in the 
exchange-correlation potential for Si obtained \cite{godby} from a GW
calculation.

Having obtained accurate results for Si, we have performed similar calculations
for Ge, the III-V zinc-blende semiconductors GaP, GaAs, GaSb, InP, InAs and InSb,
two zinc-blende oxides MgO and CaO, and one Wurzite structure oxide ZnO. The
results are displayed in Table 1 and Fig. 3.  It is evident from the results that our
procedure yields uniformly accurate results for the band gaps of narrow band-gap
to large band-gap materials.  However, for semiconductors 
like InAs and InSb which actually turn out to be metals with zero Kohn-Sham gaps, our 
 procedure is problematic. Therefore we did the initial calculation with a 
lattice constant slightly smaller than the equilibrium one, so that a very small
Kohn-Sham gap appears and then redid our calculations. The estimated band gaps in the MLSDA are slight over-estimations.  

\begin{table}[th!]
\centering
\caption{\label{tab1} Comparison of the calculated band gap ($\Delta$E) by the present work (LMTO-MLDA)
 with the TB-LMTO Kohn-Sham gap, the LMTO based Hartree-Fock (LMTO-HF) gap, the LMTO based exact exchange gap (LMTO-EXX)
 and the experimental (Exp) gap for diamond lattice Si and Ge, zinc-blende III-V semiconductors and
wurzite oxides of Zn, Ca and Mg.} 
\begin{tabular}{lcccccc}
\hline
\hline
System & a & LMTO & LMTO &LMTO&LMTO&Exp   \\
&  & (LDA) & (MLDA) & (HF) & (EXX)& \\
        & $\ang$  & eV& eV&eV & eV & eV   \\ 
\hline
\hline
Si &5.43\ \cite{cox5} &0.49 & 1.01 & 5.6\ \cite{cox} & 1.25\ \cite{kotsi} & 1.17\ \cite{cox2}  \\
Ge   &5.55& 0.08 & 0.53 & 4.2\ \cite{cox} & 1.57\ \cite{kotani1} & 0.74\ \cite{kit}  \\
GaP & 5.45\ \cite{kit}&1.62 & 2.30 &-& - & 2.32\ \cite{kit} \\
GaAs & 5.65\ \cite{cox5}&0.37 & 1.59 & -& - & 1.52\ \cite{kit}  \\
GaSb & 6.00\ \cite{cox5}&0.07 & 0.94 & -&- & 0.81\ \cite{kit}  \\
InP & 5.87\ \cite{cox5}&0.71 & 1.65 &-& - & 1.42\ \cite{kit} \\
InAs & 6.04\ \cite{cox5}&0.03 & 0.61 & - & - & 0.43\ \cite{kit}  \\
InSb &6.48\ \cite{cox5} & 0.01 & 0.59& - & - & 0.23\ \cite{kit}  \\
MgO &4.21\ \cite{cox7}& 4.49 & 7.64 &25.30\ \cite{coxa}& 7.77\ \cite{cox3} & 7.83\ \cite{cox4} \\
CaO &4.80\ \cite{cox7}& 3.35 & 6.85 &15.80\ \cite{coxa}& 7.72\ \cite{cox3} & 7.09\ \cite{cox4} \\
ZnO &3.25,5.21\ \cite{cox6}&0.87 & 3.35 & -&- & 3.44\ \cite{kit} \\
\hline\hline
\end{tabular}
\end{table}

 As mentioned before, since the band gap is calculated as an energy difference, we expect the difference in self-energy 
 correction (SIC) between the added
and the removed states to be small. We have carried out SIC corrections,
 as proposed by Samal and Harbola \cite{sh1,sh2}, for Si, GaSb and ZnO.  For Si the gap changes from 1.01 eV to 1.02 eV; for GaSb
from 0.94 eV to 1.06 eV and for ZnO from 3.35 eV to 3.37 eV. In all cases the correction is $
\sim$ 1\% confirmig our assumption.

Our correction is to the exchange
energy functional alone. Corrections to the correlation part is being implemented in a subsequent communication. However, even with the pure exchange correction major part of the correction to the Kohn-Sham gap is achieved. We are now applying this technique to a series of semi-conducting and insulating systems.

\section*{Acknowledgments}
AM would like to thank the Department of Physics, IIT, Kanpur for hosting
him during his sabbatical when this work was completed. MR and SG would like to thank the IIT, Kanpur for 
providing its academic and computational infrastructure during the time this work was carried out. We would like to thank Prof. N. Satyamurthy, IIT, Kanpur for his critical
comments on the draft of this work.

\end{document}